\pdfoutput=1
\documentclass[a4paper,fleqn]{cas-sc}
\usepackage[authoryear]{natbib}
\usepackage[utf8]{inputenc}
\usepackage[nomath]{lmodern}
\usepackage{textcomp}
\usepackage{adjustbox}
\usepackage{float}
\newsavebox{\EAAItablebox}
\ExplSyntaxOn
\cs_set:Npn \__first_footerline: { \mbox{} }
\NewDocumentCommand\EAAITable{m m +m}{%
  \sbox{\EAAItablebox}{\ignorespaces#3\unskip}%
  \dim_set:Nn \l_tbl_width_dim {\wd\EAAItablebox}
  \caption{#1}\label{#2}%
  \par\noindent\usebox{\EAAItablebox}\par}
\ExplSyntaxOff
\newcommand{\EAAIkeepwith}[1]{%
  \par\begingroup\dimen0=#1\relax
  \ifdim\dimexpr\pagegoal-\pagetotal\relax<\dimen0\newpage\fi
  \endgroup}
\hypersetup{pdfauthor={},pdftitle={}}
\begin{document}
\let\WriteBookmarks\relax
\let\printorcid\relax
\shorttitle{}
\title[mode=title]{RPL-UIE: Reference-Guided Prior Learning for Content-Preserving Underwater Image Enhancement}
\author[1,2]{Yifan Chen}
\author[2]{Jijun Lu}
\author[2]{Ye Zheng}
\author[3]{Kaichen Chi}
\author[2,3]{Zhe Sun}
\cormark[1]
\author[1]{Tao Chen}
\cormark[1]

\affiliation[1]{organization={College of Future Information Technology, Fudan University},
  city={Shanghai}, postcode={200433}, country={China}}
\affiliation[2]{organization={Institute of Artificial Intelligence (TeleAI), China Telecom},
  country={China}}
\affiliation[3]{organization={School of Artificial Intelligence, Optics and Electronics (iOPEN), Northwestern Polytechnical University},
  city={Xi'an}, postcode={710072}, country={China}}

\cortext[1]{Corresponding authors: Zhe Sun and Tao Chen.}
\nonumnote{This work is supported by National Key R\&D Program of China (No. 2026YFE0101200).}

\begin{abstract}
Underwater image enhancement (UIE) aims to recover clear images from observations degraded by wavelength-dependent absorption, scattering, and spatially nonuniform degradation. Although existing generative methods can handle complex degradations, they may still generate visually appealing results while compromising content preservation, thereby degrading downstream visual perception. To improve content preservation, we extract spatial priors characterizing appearance and photometric properties from ground-truth images to provide stable guidance for enhancement. To bridge the gap between training with ground-truth references and inference without references, we propose reference-guided prior learning for underwater image enhancement (RPL-UIE), a two-stage teacher-student framework. The teacher network extracts stable priors from ground-truth images, while the student network takes only degraded images as input and learns to mimic the teacher's prior extraction capability. We further design Residual Prior Refinement Diffusion (RPRD) and Frequency-Aware Prior Residual Calibration (FPRC) to reduce the prior discrepancy between the teacher and student. RPRD progressively predicts corrections in the residual space, while FPRC preserves stable low-frequency residuals and selectively modulates high-frequency details. Experiments on multiple underwater image datasets show that RPL-UIE achieves competitive enhancement performance. For the engineering application of underwater visual perception, downstream object detection and instance segmentation experiments show that RPL-UIE improves visual quality while better preserving object structures and semantic content needed for reliable perception. Tests on real-world data collected by a custom-built remotely operated vehicle (ROV) further demonstrate its effectiveness and practical applicability.
\end{abstract}


\maketitle
\section{Introduction}
Underwater image enhancement (UIE) is fundamental to reliable underwater visual perception, which supports applications such as marine observation, underwater robotics, object detection, and resource exploration~\citep{UCV-Survey,WROI}. Beyond conventional camera-based perception, learning-driven computational imaging has also been explored to recover visual information under scattering and underwater conditions~\citep{yf}. However, wavelength-dependent absorption, scattering, and insufficient illumination often degrade underwater images, resulting in pronounced color casts, low contrast, blurred details, and reduced visibility~\citep{Revised-UIFM,Sea-Thru,ExtremeDepthWROI}. In particular, long-wavelength light, especially red light, attenuates more rapidly in water, giving underwater images a predominant blue-green cast. Suspended particles further induce scattering, leading to localized haze and structural degradation. Moreover, spatial variations in illumination, scene depth, and water properties cause degradation to vary substantially across regions within a single image~\citep{SSDA-UIE}. Effective UIE therefore requires not only global color and brightness recovery but also degradation-aware guidance that helps preserve scene structures and visual content.

In recent years, advances in deep learning have driven UIE research from traditional enhancement approaches toward learning-based restoration frameworks. Traditional methods primarily rely on physical imaging models, color correction, Retinex theory, image fusion, or handcrafted degradation priors to mitigate color casts and visibility loss~\citep{cbf,HLRP}. However, their dependence on simplified models and handcrafted assumptions can limit their adaptability to complex underwater conditions. Learning-based methods built upon CNNs, Transformers, and generative models have been widely applied to UIE, achieving substantial progress in color correction, structural recovery, and complex degradation modeling~\citep{U-shape,SFGNet,DiffUIE,DiffColor}. Related computational-imaging studies have further explored learning-based reconstruction under complex conditions~\citep{aegi,gilm}. More recently, frequency-domain information and degradation representations have also been incorporated into conditional modeling to better characterize complex underwater degradation~\citep{WF-Diff,DiffColor,DCD-UIE}.

Despite these advances, content preservation remains challenging in generative UIE, particularly under severe underwater degradation. Existing prior-guided methods often derive priors related to color correction, physical cues, or degradation representations from degraded observations and use them as auxiliary conditions during enhancement~\citep{Ucolor,GuidedHybSensUIR,DiffUIE,SeaDiff,PA-Diff}. However, because these conditional representations are inferred from low-quality inputs, their accuracy may be affected by color casts, scattering-induced blur, and local information loss. Consequently, they may fail to faithfully characterize the underlying scene, and the resulting representation errors can propagate through the generative restoration process, potentially causing structural or semantic content distortion in the restored images.

To address this issue, we propose RPL-UIE, a two-stage teacher--student framework for reference-guided prior learning. In Stage I, the teacher extracts clean appearance and photometric information from ground-truth images while using the corresponding degraded observations to retain input-specific context. The resulting spatial priors guide image reconstruction and serve as supervision targets for the student stage. In Stage II, the student network takes only the degraded image as input and learns to estimate the corresponding coarse priors under teacher-prior supervision. This enables the student to approximate the reference-guided teacher priors using only degraded images at inference.

Nevertheless, priors estimated by the student may still exhibit discrepancies from their teacher counterparts because of the information loss in degraded observations. To reduce this teacher--student prior discrepancy, we introduce Residual Prior Refinement Diffusion (RPRD), which uses the student-estimated coarse priors as anchors and progressively predicts the necessary residual corrections toward the teacher priors. Rather than directly predicting complete teacher priors, this residual formulation preserves informative components already captured by the coarse priors while focusing on the required corrections. We further develop Frequency-Aware Prior Residual Calibration (FPRC), which accounts for the unequal reliability of different residual components. FPRC retains stable low-frequency residual components and selectively modulates less reliable high-frequency detail residuals according to their spatial-channel reliability, producing calibrated priors for image reconstruction. Together, these designs provide reference-guided priors for content-preserving enhancement without requiring reference images at inference.

The main contributions of this work are summarized as follows:
\begin{itemize}

    \item We propose RPL-UIE, a two-stage teacher--student framework for reference-guided prior learning. The teacher learns complementary appearance and photometric priors from ground-truth information, while the student learns to estimate the corresponding priors from degraded inputs for content-preserving enhancement at inference.
    
    \item We introduce Residual Prior Refinement Diffusion (RPRD), which reduces the teacher--student prior discrepancy by modeling the correction from student coarse priors to teacher priors through residual-space diffusion.

    \item We develop Frequency-Aware Prior Residual Calibration (FPRC), which preserves stable low-frequency corrections while adaptively modulating high-frequency details to obtain calibrated spatial priors.
    
     \item Extensive experiments on multiple UIE benchmarks demonstrate competitive enhancement performance. Downstream detection and instance segmentation further support the utility of content-preserving enhancement for visual perception, while real-world ROV evaluations support its practical applicability.
\end{itemize}

\section{Related Work}

\subsection{Traditional Underwater Image Enhancement}

Early underwater image enhancement methods mainly rely on physical image-formation models, color correction, Retinex theory, image fusion, or handcrafted degradation priors. Physics-based approaches estimate background light, transmission, scene depth, or wavelength-dependent attenuation to restore image color and visibility. Chiang and Chen combine a dehazing model with wavelength compensation while accounting for artificial illumination, whereas the red-channel prior extends the dark-channel concept by exploiting the rapid attenuation of red light underwater~\citep{WCID,Red-Channel-prior}.
Beyond explicit model inversion, color balancing, Retinex reflectance priors, contrast enhancement, and multiscale fusion have also been explored. Ancuti et al. combine white balancing with multiscale fusion to improve color and local contrast~\citep{cbf}. Zhuang et al. incorporate Retinex theory and a hyper-Laplacian reflectance prior to recover edges, details, and color naturalness~\citep{HLRP}. 

These methods are highly interpretable and have provided useful insights for subsequent learning-based UIE methods, particularly in color compensation, structural enhancement, and frequency decomposition. However, handcrafted priors, empirical parameters, and simplified imaging assumptions often struggle to accommodate complex underwater conditions, nonuniform illumination, and severe color casts. Even with local adaptation, multiple priors, or fusion strategies, their restoration performance may remain sensitive to scene variations, making it difficult to consistently balance color naturalness, local contrast, and texture preservation.

\subsection{Deep Learning-Based Underwater Image Enhancement}

Deep learning has advanced UIE by learning end-to-end restoration mappings from degraded images to clear images. UIEB introduced a real-world benchmark and Water-Net, which fuses white-balanced, histogram-equalized, and gamma-corrected inputs, while NPT-UL uses nonphysical transformations and unsupervised learning for underwater image enhancement~\citep{UIEB,NPT-UL}. Subsequent networks introduce color-space transformations, physical cues, and multiscale or contextual modeling. UColor combines multi-color-space encoding with medium-transmission guidance, whereas U-shape introduces channel-wise multiscale feature fusion and spatial-wise global modeling within a U-shaped Transformer~\citep{Ucolor,U-shape}.
More recent methods explore staged constraints, prior-guided restoration, and spatial--frequency interaction. CCL-Net uses cascaded contrastive learning for color correction and dehazing; GuidedHybSensUIR combines local restoration, contextual modeling, and a color-balance prior; and SFGNet enhances spatial--frequency representations~\citep{CCL-Net,GuidedHybSensUIR,SFGNet}. FDCE-Net, TAFormer, and WWE-UIE further incorporate frequency-domain degradation decoupling, transmission-aware modeling, and adaptive white balancing with wavelet and gradient-aware enhancement, respectively~\citep{FDCE-Net,TAFormer,WWE-UIE}.

Overall, deep learning-based UIE methods have advanced color correction, structural recovery, and adaptation to complex scenes. However, simultaneous color shifts, nonuniform illumination, scattering-induced blur, and texture loss continue to challenge cross-scene generalization and content preservation.
\begin{figure}[pos=!t]
\centering
\includegraphics[
    width=\textwidth,
    trim={0 36bp 0 0},
    clip
]{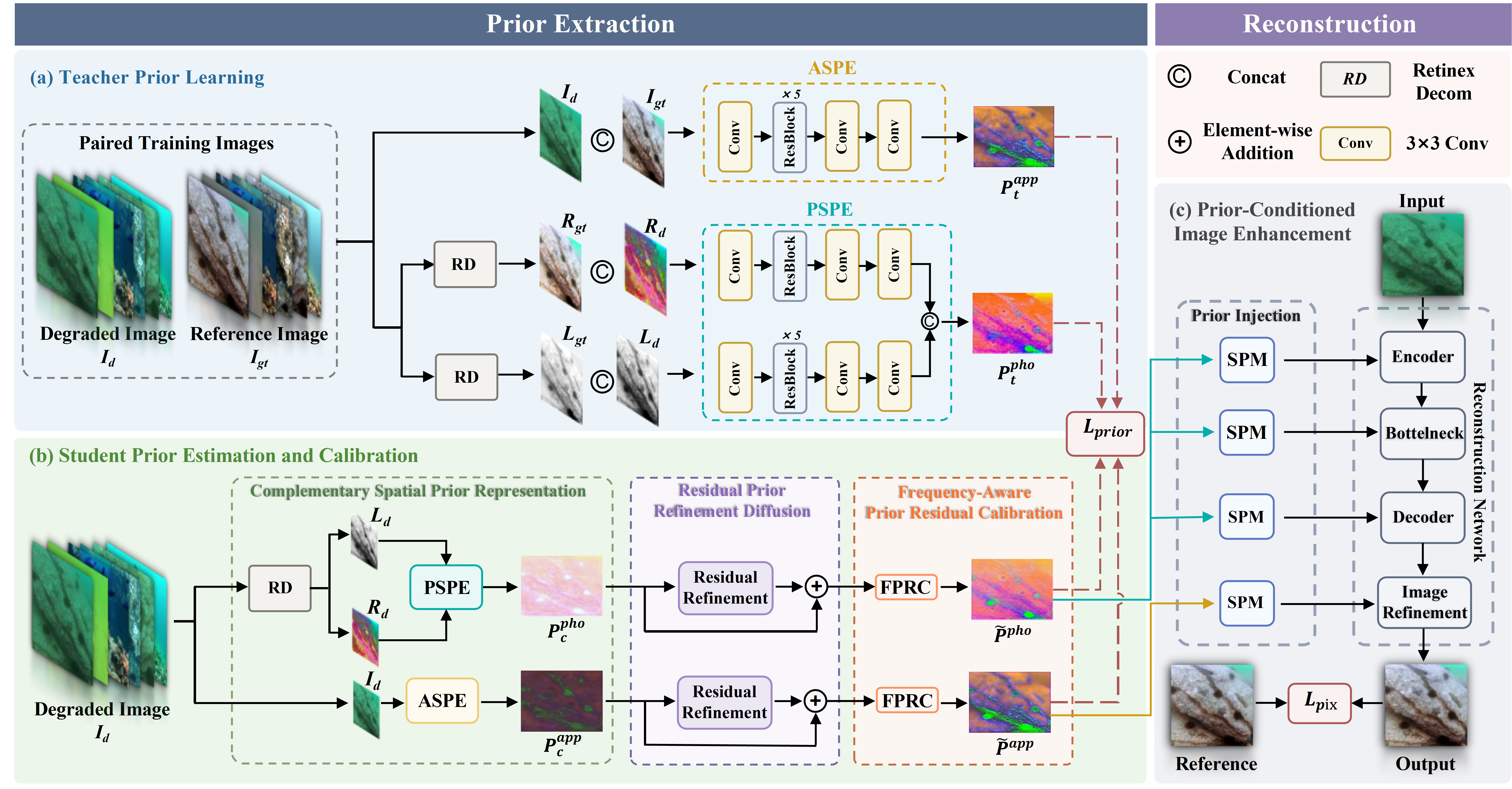}
\caption{Overall framework of RPL-UIE.
(a) In Stage I, the Appearance Spatial Prior Encoder (ASPE) and Photometric Spatial Prior Encoder (PSPE) learn reference-guided appearance and photometric teacher priors from ground-truth information while incorporating context from the corresponding degraded observations. (b) In Stage II, the student pathway estimates the corresponding coarse priors from the degraded input alone. Under teacher-prior supervision, RPRD progressively refines the coarse priors toward their teacher counterparts, while FPRC calibrates the resulting prior residuals to obtain the final calibrated priors. (c) The calibrated priors obtained in Stage II are injected into the reconstruction network through Spatial Prior Modulation (SPM) to guide image enhancement.}
\label{fig:pipeline}
\end{figure}

\subsection{Diffusion and Generative Models for Underwater Image Enhancement}

In recent years, diffusion models have been adopted for conditional restoration in UIE. Pixel-space methods perform progressive noising and denoising in RGB space while conditioning restoration on degraded images or additional cues. DCGF introduces diffusion-color guidance for underwater enhancement, whereas SU-DDPM accelerates conditional DDPM sampling~\citep{DCGF,SU-DDPM}. PA-Diff, W2WDiff, Water-CDM, UADiff, and DCD-UIE further introduce physical priors, unsupervised water-to-water conversion, Retinex-based brightness correction, uncertainty-aware learning, and perceptual color--structure decoupling, respectively~\citep{PA-Diff,W2WDiff,Water-CDM,UADiff,DCD-UIE}.
Several methods instead introduce latent or feature priors to reduce modeling difficulty and provide more compact restoration guidance. DiffUIE learns a global feature prior, Reti-Diff generates reflectance and illumination priors through latent diffusion, and SLURPP combines a pretrained latent diffusion model with explicit scene decomposition~\citep{DiffUIE,Reti-Diff,SLURPP}. Frequency- and wavelet-domain methods model transformed low-frequency information or frequency residuals for color correction and detail recovery. WF-Diff refines an initial restoration through frequency residual diffusion, whereas DiffColor conducts diffusion in the wavelet low-frequency domain with high-frequency detail compensation~\citep{WF-Diff,DiffColor}.

Nevertheless, existing diffusion-based UIE methods mainly focus on conditioning design, latent-space modeling, or frequency-domain transformations, while priors inferred solely from degraded observations may provide insufficient guidance for content preservation. In contrast, RPL-UIE exploits ground-truth information during training to learn reference-guided priors while retaining reference-free inference.

\section{Method}
\subsection{Overview}

Given a degraded underwater image $\mathbf{I}_{d}$, UIE aims to recover an enhanced image $\hat{\mathbf{I}}$ with natural colors, well-balanced contrast, and clear structural details. As illustrated in Fig.~\ref{fig:pipeline}, RPL-UIE adopts a two-stage teacher--student framework comprising teacher prior learning, student prior estimation and calibration, and prior-conditioned image enhancement.

In Stage I, the Appearance Spatial Prior Encoder (ASPE) and Photometric Spatial Prior Encoder (PSPE) learn complementary reference-guided teacher spatial priors using ground-truth images together with the corresponding degraded observations. The ground-truth images provide clear-scene appearance and reflectance--illumination information, while the degraded observations provide input-specific degradation context.
These teacher priors guide image reconstruction in Stage I and provide supervision targets for prior learning in Stage II.

In Stage II, ASPE and PSPE estimate the corresponding appearance and photometric coarse priors solely from the degraded input. 
During training, the teacher priors define branch-wise residual targets for RPRD, which progressively refines each coarse prior toward its teacher counterpart. FPRC then calibrates the refinement residual by retaining low-frequency components and adaptively modulating high-frequency details according to their reliability.
The resulting calibrated priors are injected into the reconstruction network to generate $\hat{\mathbf{I}}$. At inference, the teacher pathway and reference image are removed, and the entire enhancement process takes only $\mathbf{I}_{d}$ as input.

\subsection{Complementary Appearance and Photometric Spatial Priors}
Underwater images retain appearance cues such as color distribution, local contrast, and texture visibility, but these cues are entangled with illumination and attenuation effects. A fixed Retinex decomposition provides complementary photometric cues by separating reflectance- and illumination-related components. Accordingly, RPL-UIE constructs complementary appearance and photometric spatial prior branches. The subscripts $t$ and $s$ denote the teacher and student prior estimators, respectively.

\subsubsection{Reference-Guided Teacher Prior Learning}
In Stage I, RPL-UIE learns complementary teacher spatial priors under ground-truth guidance. Given a degraded training image $\mathbf{I}_{d}$ and its corresponding ground-truth image $\mathbf{I}_{gt}$, the Appearance Spatial Prior Encoder (ASPE) jointly processes both images. The ground-truth image provides clean appearance information, while the degraded image supplies observation-specific context. The appearance teacher prior is estimated as
\begin{equation}
\mathbf{P}^{\mathrm{app}}_{t}
=
\operatorname{ASPE}_{t}
\left(
\mathbf{I}_{d},
\mathbf{I}_{gt}
\right).
\label{eq:appearance_teacher_prior}
\end{equation}
The resulting $\mathbf{P}^{\mathrm{app}}_{t}$ characterizes clear-scene appearance properties, including color distribution, contrast, and texture visibility.

To construct the complementary photometric prior, fixed Retinex decomposition operators first decompose the degraded and ground-truth images into reflectance- and illumination-related components:
\begin{equation}
(\mathbf{R}_{d},\mathbf{L}_{d})
=
\mathcal{D}_{l}(\mathbf{I}_{d}),
\label{eq:photometric_decomposition_lq}
\end{equation}
\begin{equation}
(\mathbf{R}_{gt},\mathbf{L}_{gt})
=
\mathcal{D}_{h}(\mathbf{I}_{gt}),
\label{eq:photometric_decomposition_gt}
\end{equation}
where $\mathbf{R}_{d}$ and $\mathbf{L}_{d}$ denote the reflectance- and illumination-related components of the degraded image, while $\mathbf{R}_{gt}$ and $\mathbf{L}_{gt}$ denote those of the ground-truth image. The ground-truth components provide clean photometric information, whereas the degraded components retain the observed degradation context. The Photometric Spatial Prior Encoder (PSPE) jointly processes these representations:

\begin{equation}
\mathbf{P}^{\mathrm{pho}}_{t}
=
\operatorname{PSPE}_{t}
\left(
\operatorname{Cat}(\mathbf{R}_{gt},\mathbf{R}_{d}),
\operatorname{Cat}(\mathbf{L}_{gt},\mathbf{L}_{d})
\right),
\label{eq:photometric_teacher_prior}
\end{equation}
where $\operatorname{Cat}(\cdot)$ denotes channel-wise concatenation. 
The resulting $\mathbf{P}^{\mathrm{pho}}_{t}$ characterizes reference-guided photometric properties associated with color attenuation, illumination variation, and reflectance visibility.

The complementary teacher spatial priors are collected as
\begin{equation}
\mathcal{P}_{t}
=
\{\mathbf{P}^{\mathrm{app}}_{t},\mathbf{P}^{\mathrm{pho}}_{t}\}
=
\{\mathbf{P}^{b}_{t}\mid b\in\{\mathrm{app},\mathrm{pho}\}\}.
\label{eq:teacher_prior_set}
\end{equation}
These teacher priors guide image reconstruction in Stage I.

\begin{figure}[pos=!t]
\centering
\includegraphics[width=\textwidth]{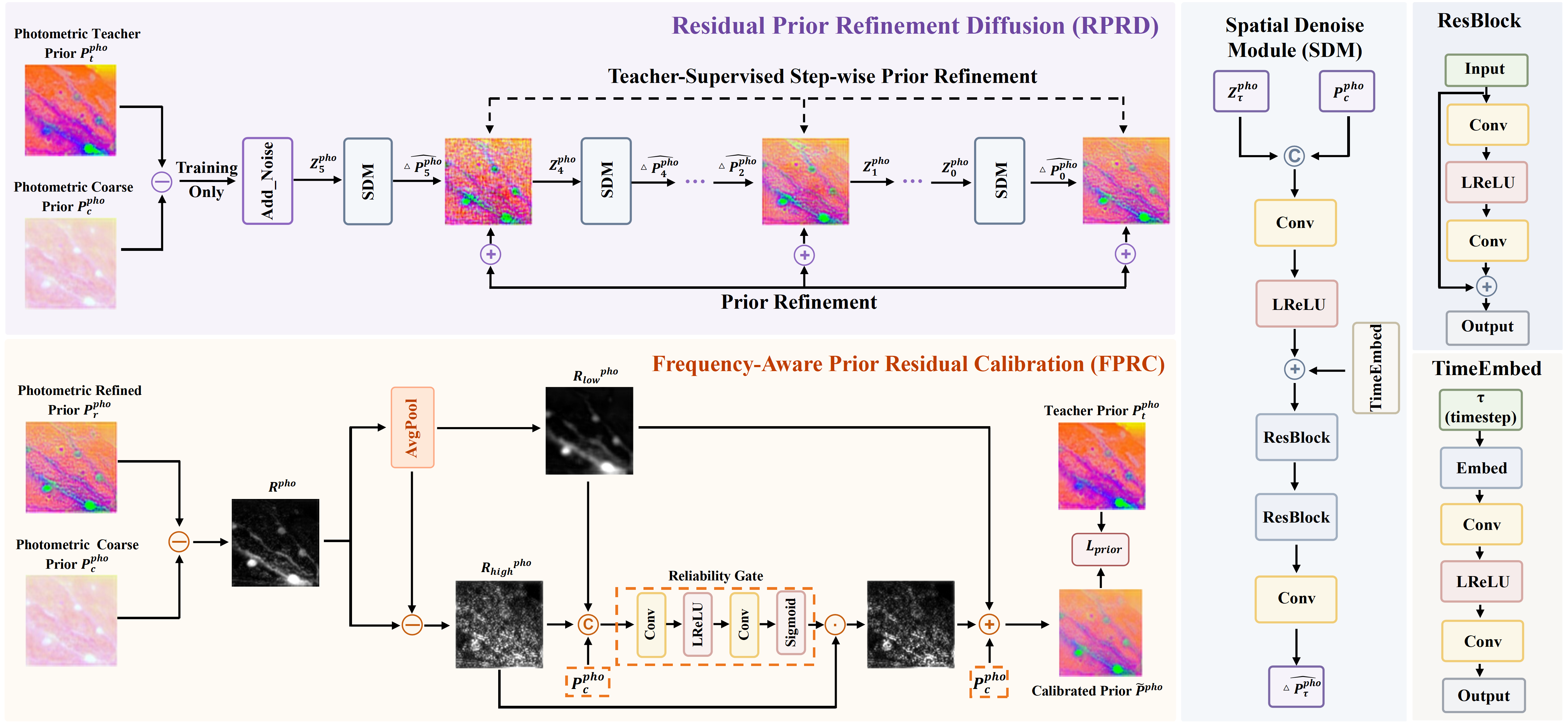}
\caption{Detailed architectures of Residual Prior Refinement Diffusion (RPRD) and Frequency-Aware Prior Residual Calibration (FPRC), illustrated with the photometric prior branch. RPRD progressively predicts residual corrections to obtain refined priors, while FPRC decomposes the corresponding prior residuals and selectively calibrates their high-frequency components to generate calibrated priors. During training, the refined prior from each reverse step is processed by FPRC, and the resulting calibrated prior is supervised by the corresponding teacher prior. The same refinement and calibration procedure is applied to the appearance prior branch.}
\label{fig:rprd_fprc_module}
\end{figure}

\subsubsection{Student Coarse Prior Estimation from Degraded Inputs}

In Stage II, the student prior encoders receive only the degraded image as input, while the frozen teacher provides reference-guided prior targets during training. ASPE first estimates the appearance spatial coarse prior:
\begin{equation}
\mathbf{P}^{\mathrm{app}}_{c}
=
\operatorname{ASPE}_{s}(\mathbf{I}_{d}),
\label{eq:appearance_coarse_prior}
\end{equation}
where $\mathbf{P}^{\mathrm{app}}_{c}$ primarily characterizes the observed color distribution, local contrast, and texture visibility of the degraded image.

Using the degraded photometric representation obtained in Eq.~\eqref{eq:photometric_decomposition_lq}, PSPE estimates the corresponding photometric spatial coarse prior:
\begin{equation}
\mathbf{P}^{\mathrm{pho}}_{c}
=
\operatorname{PSPE}_{s}
\left(
\operatorname{Cat}(\mathbf{R}_{d},\mathbf{L}_{d})
\right),
\label{eq:photometric_coarse_prior}
\end{equation}
where $\mathbf{P}^{\mathrm{pho}}_{c}$ emphasizes spatial variations in color attenuation, illumination, and reflectance visibility, thereby complementing the appearance prior.

The complementary spatial coarse priors are collected as
\begin{equation}
\mathcal{P}_{c}
=
\{\mathbf{P}^{\mathrm{app}}_{c},\mathbf{P}^{\mathrm{pho}}_{c}\}
=
\{\mathbf{P}^{b}_{c}\mid b\in\{\mathrm{app},\mathrm{pho}\}\}.
\label{eq:coarse_prior_set}
\end{equation}
During training, the corresponding teacher priors $\mathcal{P}_{t}$ define branch-wise residual targets and supervise the subsequent refinement and calibration of the coarse priors.

\subsection{Residual Prior Refinement Diffusion}

For either spatial prior branch, $\mathbf{P}^{b}_{c}$ provides an initial conditional representation for image reconstruction, where $b\in\{\mathrm{app},\mathrm{pho}\}$ denotes the appearance or photometric prior branch. Because $\mathbf{P}^{b}_{c}$ is estimated solely from the degraded input, its accuracy may be affected in regions with severe color distortion, low visibility, or strong scattering. Directly using the coarse prior may therefore propagate representation errors into image reconstruction. Meanwhile, directly predicting the complete teacher prior may increase the learning difficulty and overwrite informative cues already captured by the coarse prior. We therefore propose Residual Prior Refinement Diffusion (RPRD), which uses the coarse prior as an anchor and learns only the residual correction toward its teacher counterpart.

For either prior branch $b$, the residual target and the corresponding refined prior are defined as
\begin{equation}
\begin{aligned}
\Delta \mathbf{P}^{b}
&=
\mathbf{P}^{b}_{t}
-
\mathbf{P}^{b}_{c}, \\
\mathbf{P}^{b}_{r}
&=
\mathbf{P}^{b}_{c}
+
\widehat{\Delta \mathbf{P}^{b}}.
\end{aligned}
\label{eq:rprd_residual_target}
\end{equation}
This residual formulation preserves informative components already captured by the coarse prior while focusing the model on the necessary corrections.

To stabilize residual learning, RPRD applies a forward diffusion process and reverse denoising in the prior residual space, following the standard diffusion formulation~\citep{DDPM}. Given $\Delta \mathbf{P}^{b}$, its noisy state at diffusion step $\tau$ is expressed as
\begin{equation}
\mathbf{Z}^{b}_{\tau}
=
\sqrt{\bar{\alpha}_{\tau}}\Delta \mathbf{P}^{b}
+
\sqrt{1-\bar{\alpha}_{\tau}}\boldsymbol{\epsilon},
\quad
\boldsymbol{\epsilon}\sim\mathcal{N}(\mathbf{0},\mathbf{I}),
\label{eq:rprd_forward_sample}
\end{equation}
where $\bar{\alpha}_{\tau}=\prod_{i=1}^{\tau}\alpha_i$, $\alpha_i=1-\beta_i$, and $\{\beta_i\}$ denotes a predefined noise schedule. RPRD operates in the residual space anchored by $\mathbf{P}^{b}_{c}$ rather than directly modeling a complete image or teacher prior.

As shown in Fig.~\ref{fig:rprd_fprc_module}, the reverse refinement process is implemented by the Spatial Denoise Module (SDM). At each step, $\operatorname{SDM}^{b}_{\phi}$ predicts the clean residual correction from the current residual state, the coarse prior, and the diffusion step:
\begin{equation}
\begin{aligned}
\widehat{\Delta \mathbf{P}^{b}_{\tau}}
&=
\operatorname{SDM}^{b}_{\phi}
\left(
\mathbf{Z}^{b}_{\tau},
\mathbf{P}^{b}_{c},
\tau
\right), \\
\mathbf{P}^{b}_{r,\tau}
&=
\mathbf{P}^{b}_{c}
+
\widehat{\Delta \mathbf{P}^{b}_{\tau}}.
\end{aligned}
\label{eq:rprd_prediction}
\end{equation}
Here, $\mathbf{P}^{b}_{c}$ provides input-dependent conditioning, while $\tau$ is encoded through a time-step embedding. SDM fuses the coarse prior with the noisy residual state and injects the time-step embedding into its residual blocks.

The residual state is then updated deterministically using the diffusion posterior mean:
\begin{equation}
\mathbf{Z}^{b}_{\tau-1}
=
\tilde{\boldsymbol{\mu}}_{\tau}
\left(
\mathbf{Z}^{b}_{\tau},
\widehat{\Delta \mathbf{P}^{b}_{\tau}}
\right),
\quad
\tau=T,\ldots,1,
\label{eq:rprd_reverse_update}
\end{equation}
where $\tilde{\boldsymbol{\mu}}_{\tau}(\cdot)$ denotes the posterior mean determined by the current residual state and the predicted clean residual correction. No additional stochastic sampling is introduced during inference.

During training, RPRD generates refined priors $\mathbf{P}^{b}_{r,\tau}$ at multiple reverse steps. At inference, RPRD starts from an all-zero residual state, $\mathbf{Z}^{b}_{T}=\mathbf{0}$, and performs a fixed number of deterministic reverse updates conditioned only on the coarse prior $\mathbf{P}^{b}_{c}$. The final refined priors of the appearance and photometric branches are denoted by $\mathbf{P}^{\mathrm{app}}_{r}$ and $\mathbf{P}^{\mathrm{pho}}_{r}$, respectively.

\subsection{Frequency-Aware Prior Residual Calibration}

The refined spatial prior $\mathbf{P}^{b}_{r}$ provides a residual correction to the spatial coarse prior $\mathbf{P}^{b}_{c}$. 
However, this correction may combine smooth low-frequency compensation with high-frequency local responses that are more susceptible to residual estimation errors. Directly applying the entire correction may therefore propagate such errors into subsequent image reconstruction. We propose Frequency-Aware Prior Residual Calibration (FPRC), which decomposes the prior residual into low- and high-frequency components and adaptively modulates the high-frequency component to produce a calibrated conditional prior.

For either prior branch $b\in\{\mathrm{app},\mathrm{pho}\}$, we first compute the residual between the refined and coarse priors:
\begin{equation}
\mathbf{R}^{b}
=
\mathbf{P}^{b}_{r}
-
\mathbf{P}^{b}_{c},
\label{eq:fprc_residual}
\end{equation}
where $\mathbf{R}^{b}$ denotes the prior residual of the refined prior relative to the coarse prior. FPRC then decomposes it into low- and high-frequency residuals using a low-pass operator $\mathcal{L}(\cdot)$:
\begin{equation}
\mathbf{R}^{b}_{\mathrm{low}}
=
\mathcal{L}(\mathbf{R}^{b}),
\quad
\mathbf{R}^{b}_{\mathrm{high}}
=
\mathbf{R}^{b}
-
\mathbf{R}^{b}_{\mathrm{low}},
\label{eq:fprc_low_high}
\end{equation}
where $\mathbf{R}^{b}_{\mathrm{low}}$ and $\mathbf{R}^{b}_{\mathrm{high}}$ denote the low- and high-frequency prior residuals, respectively. We implement $\mathcal{L}(\cdot)$ using average pooling to obtain a smooth low-frequency residual estimate without introducing additional complex transforms.

The low-frequency residual primarily captures region-level compensation associated with color, illumination, and overall degradation trends, providing a smooth correction to the coarse prior. By contrast, the high-frequency residual contains more localized detail responses and is more susceptible to residual estimation errors. FPRC therefore learns a spatial-channel reliability gate $\mathbf{M}^{b}$ to adaptively modulate the high-frequency detail residual:
\begin{equation}
\mathbf{M}^{b}
=
\sigma
\left(
\mathcal{H}^{b}
\left(
\operatorname{Cat}
\left(
\mathbf{P}^{b}_{c},
\mathbf{R}^{b}_{\mathrm{low}},
\mathbf{R}^{b}_{\mathrm{high}}
\right)
\right)
\right),
\label{eq:fprc_gate}
\end{equation}
where $\mathcal{H}^{b}(\cdot)$ is the lightweight gate-prediction network shown in Fig.~\ref{fig:rprd_fprc_module}, comprising two convolutional layers and a LeakyReLU activation, and $\sigma(\cdot)$ is the sigmoid function. The resulting gate $\mathbf{M}^{b}$ estimates the reliability of high-frequency residual responses at each spatial location and channel.

Finally, FPRC combines the low-frequency residual and the gated high-frequency residual with the coarse prior to obtain the calibrated spatial prior:
\begin{equation}
\tilde{\mathbf{P}}^{b}
=
\mathbf{P}^{b}_{c}
+
\mathbf{R}^{b}_{\mathrm{low}}
+
\mathbf{M}^{b}
\ast
\mathbf{R}^{b}_{\mathrm{high}}.
\label{eq:fprc_calibrated_prior}
\end{equation}
This design retains stable low-frequency compensation while adaptively modulating high-frequency detail residuals according to their estimated reliability, thereby reducing the influence of unstable local responses on the reconstruction network.

Applying FPRC to the two prior branches yields the calibrated appearance and photometric spatial priors $\tilde{\mathbf{P}}^{\mathrm{app}}$ and $\tilde{\mathbf{P}}^{\mathrm{pho}}$, respectively.

\subsection{Prior-Conditioned Reconstruction and Training Objectives}

The reconstruction pathway is conditioned by the teacher priors in Stage I and by the calibrated priors in Stage II. As shown in Fig.~\ref{fig:pipeline}, the photometric spatial prior conditions the encoder--bottleneck--decoder reconstruction path, whereas the appearance spatial prior conditions the final image refinement stage. This branch-specific injection is implemented through Spatial Prior Modulation (SPM), which converts the corresponding spatial prior into a multiplicative conditioning signal for the reconstruction features. During Stage II, given the feature $\mathbf{F}_{l}$ at layer $l$ and the calibrated prior $\tilde{\mathbf{P}}^{b}$ from the corresponding branch, SPM is formulated as
\begin{equation}
\mathbf{F}^{\prime}_{l}
=
\mathbf{F}_{l}
\ast
\left(
1+
\tanh
\left(
\Pi_{l}(\tilde{\mathbf{P}}^{b})
\right)
\right),
\label{eq:prior_feature_modulation}
\end{equation}
where $\Pi_{l}(\cdot)$ denotes a spatial prior projector that performs scale alignment followed by a $1\times1$ convolution. Here, $b=\mathrm{pho}$ for the main reconstruction path and $b=\mathrm{app}$ for the final refinement stage. This bounded modulation adaptively amplifies or attenuates the reconstruction features according to the corresponding calibrated spatial prior. The final enhanced image is generated as
\begin{equation}
\hat{\mathbf{I}}
=
\mathcal{G}_{\theta}
\left(
\mathbf{I}_{d},
\tilde{\mathbf{P}}^{\mathrm{pho}},
\tilde{\mathbf{P}}^{\mathrm{app}}
\right),
\label{eq:prior_conditioned_reconstruction}
\end{equation}
where $\mathcal{G}_{\theta}$ denotes the prior-conditioned image reconstruction network.

Within the prior-conditioned reconstruction network, a lightweight decomposition module is reused at the shallow encoder and late decoder levels to provide complementary supervision signals. For each selected feature map, the module estimates reflectance and illumination components. At the shallow encoder level, their product forms an observation reconstruction, whereas at the late decoder level, their channel-wise concatenation forms a photometric representation.

RPL-UIE is optimized in two stages. In Stage I, the teacher pathway and reconstruction network are jointly optimized using

\begin{equation}
\mathcal{L}_{\mathrm{S1}}
=
\mathcal{L}_{\mathrm{pix}}
+
\mathcal{L}_{\mathrm{obs}}
+
\mathcal{L}_{\mathrm{pho}},
\label{eq:stage1_loss}
\end{equation}

where all terms are $L_{1}$ losses. $\mathcal{L}_{\mathrm{pix}}$ constrains the Stage I enhanced output to match $\mathbf{I}_{gt}$. $\mathcal{L}_{\mathrm{obs}}$ constrains the observation reconstruction from the shallow encoder level to match the degraded input $\mathbf{I}_{d}$, while $\mathcal{L}_{\mathrm{pho}}$ aligns the photometric representation from the late decoder level with $\operatorname{Cat}(\mathbf{R}_{gt},\mathbf{L}_{gt})$.

After Stage I pretraining, the teacher prior encoders are frozen, and the resulting teacher priors are used as fixed supervision targets for student prior learning. Stage II then supervises prior refinement and calibration at every reverse step. Let $\tilde{\mathbf{P}}^{b}_{k}$ denote the calibrated prior obtained at the $k$-th reverse step. The multi-step prior loss is defined as
\begin{equation}
\mathcal{L}_{\mathrm{prior}}
=
\sum_{b\in\{\mathrm{app},\mathrm{pho}\}}
\sum_{k=1}^{T}
\omega_{k}
\left\|
\tilde{\mathbf{P}}^{b}_{k}
-
\mathbf{P}^{b}_{t}
\right\|_{1},
\label{eq:multi_step_prior_loss}
\end{equation}
where $T$ is the number of reverse refinement steps, $\omega_{k}$ is the supervision weight for the $k$-th step, and $\mathbf{P}^{b}_{t}$ denotes the teacher prior for branch $b$. Thus, the teacher prior supervises the calibrated predictions produced at all reverse steps of the corresponding branch.

Stage II begins with a prior warm-up phase using only
$\mathcal{L}_{\mathrm{prior}}$, followed by joint optimization with
\begin{equation}
\mathcal{L}_{\mathrm{S2}}
=
\mathcal{L}_{\mathrm{pix}}
+
\lambda_{\mathrm{prior}}\mathcal{L}_{\mathrm{prior}}
+
\mathcal{L}_{\mathrm{pho}},
\label{eq:stage2_loss}
\end{equation}
where $\mathcal{L}_{\mathrm{pix}}$ and $\mathcal{L}_{\mathrm{pho}}$ follow the definitions above and are evaluated on the Stage II reconstruction outputs. Following Stage I pretraining, Stage II omits $\mathcal{L}_{\mathrm{obs}}$ and jointly optimizes image reconstruction and teacher-supervised calibrated-prior learning for both branches through $\mathcal{L}_{\mathrm{prior}}$. The coefficient $\lambda_{\mathrm{prior}}$ was set to $0.35$.

\section{Experiments}

\subsection{Experimental Settings}

\subsubsection{Datasets}
Experiments were conducted on five paired underwater datasets: UIEB~\citep{UIEB}, LSUI~\citep{U-shape}, and the underwater-dark, underwater-imagenet, and underwater-scenes subsets of EUVP~\citep{EUVP}. UIEB and LSUI contain 890 and 4,279 image pairs, respectively, whereas EUVP-underwater-dark, EUVP-underwater-imagenet, and EUVP-underwater-scenes contain 5,550, 3,700, and 2,185 image pairs, respectively. Each dataset was divided into training and test sets at a ratio of 8:2. During training, all input images were resized to \(256\times256\).

\begin{figure}[pos=!htbp]
\centering
\includegraphics[width=\textwidth]{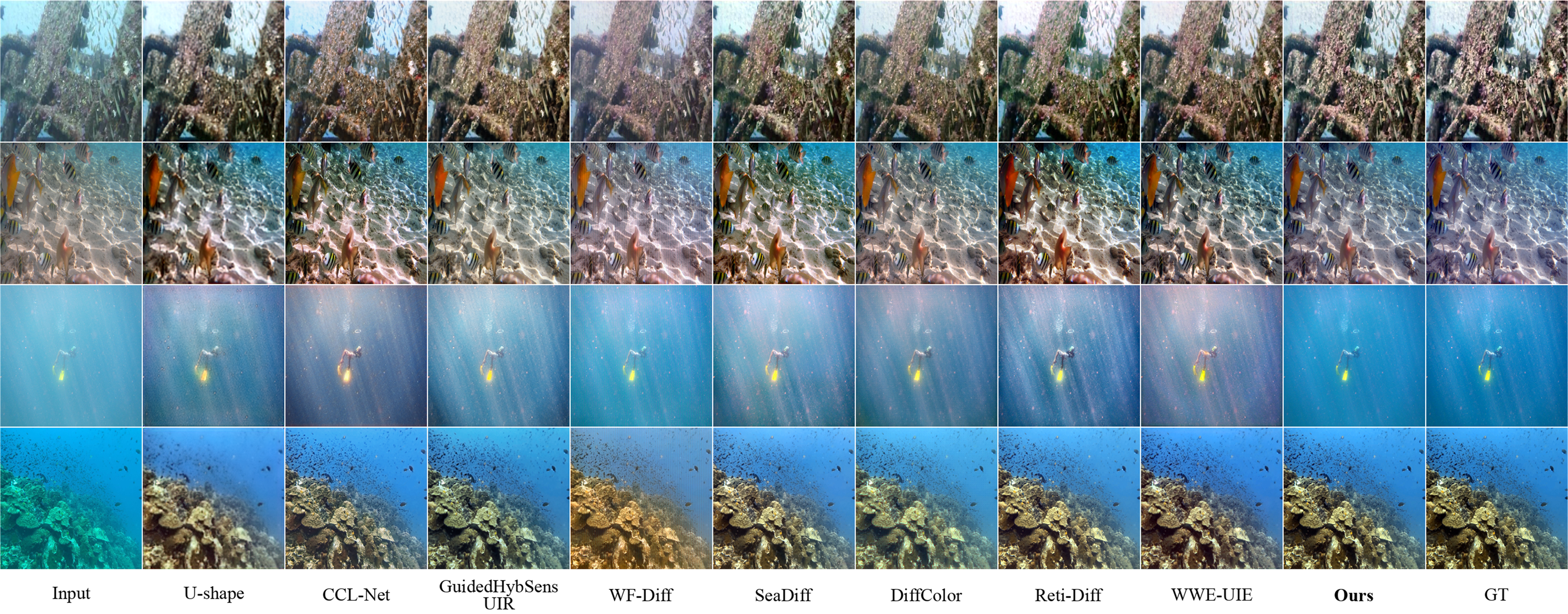}
\caption{Visual comparison on the UIEB dataset.}
\label{fig:visual_comparison}
\end{figure}

\subsubsection{Implementation Details}

RPL-UIE was implemented in PyTorch, and all experiments were performed on an NVIDIA H100 GPU. Stage I was trained for 300k iterations. Stage II was trained for 400k iterations on UIEB and 800k iterations on LSUI and the three EUVP subsets. We used the AdamW optimizer with a batch size of 2 per GPU. The initial learning rate in Stage I was \(2\times10^{-4}\) and was adjusted using cosine annealing. During Stage II, the prior-learning warm-up used an initial learning rate of \(1\times10^{-4}\). During joint optimization, the reconstruction-network and prior parameter groups used initial learning rates of \(2\times10^{-4}\) and \(5\times10^{-5}\), respectively. The learning rates were decayed every 80k iterations. The number of RPRD reverse refinement steps was set to six. 

\subsubsection{Evaluation Metrics}
We evaluated the enhanced images using PSNR, SSIM~\citep{SSIM}, LPIPS~\citep{LPIPS}, UIQM~\citep{UIQM}, and UCIQE~\citep{UCIQE}. PSNR and SSIM measure pixel-level fidelity and structural similarity between the enhanced and reference images, whereas LPIPS measures their perceptual discrepancy. UIQM and UCIQE provide no-reference quality assessments based on underwater image attributes such as color, contrast, and sharpness.

\subsection{Comparison with State-of-the-Art Methods}

\subsubsection{Compared Methods}

To comprehensively evaluate RPL-UIE, we compared it with eight representative UIE methods: U-shape~\citep{U-shape}, CCL-Net~\citep{CCL-Net}, GuidedHybSensUIR~\citep{GuidedHybSensUIR}, WF-Diff~\citep{WF-Diff}, SeaDiff~\citep{SeaDiff}, DiffColor~\citep{DiffColor}, Reti-Diff~\citep{Reti-Diff}, and WWE-UIE~\citep{WWE-UIE}.
All methods were evaluated using the same data splits and a unified evaluation protocol.

\EAAIkeepwith{0.49\textheight}
\begin{table}[pos=H]
\centering
\normalfont\rmfamily\fontsize{7.5}{8.4}\selectfont
\setlength{\tabcolsep}{2.0pt}
\renewcommand{\arraystretch}{1.03}

\EAAITable{Quantitative comparison on the UIEB and LSUI datasets. Best results are in bold.}{tab:main_results_uieb_lsui}{%
\begin{adjustbox}{max width=\linewidth}
\begin{tabular}{@{}llcccccccccc@{}}
\toprule
Method & Venue
& \multicolumn{5}{c}{UIEB}
& \multicolumn{5}{c}{LSUI} \\
\cmidrule(lr){3-7}\cmidrule(lr){8-12}
& & PSNR\textuparrow & SSIM\textuparrow & LPIPS\textdownarrow & UIQM\textuparrow & UCIQE\textuparrow
& PSNR\textuparrow & SSIM\textuparrow & LPIPS\textdownarrow & UIQM\textuparrow & UCIQE\textuparrow \\
\midrule
\makecell[l]{U-shape\\\citep{U-shape}} & TIP'23
& 21.84 & 0.804 & 0.259 & 4.268 & 0.611
& 24.76 & 0.819 & 0.201 & 4.231 & 0.586 \\
\makecell[l]{CCL-Net\\\citep{CCL-Net}} & TMM'25
& 21.76 & 0.883 & 0.152 & \textbf{4.456} & 0.614
& 24.57 & 0.839 & 0.145 & \textbf{4.340} & 0.583 \\
\makecell[l]{GuidedHybSensUIR\\\citep{GuidedHybSensUIR}} & TCSVT'25
& 23.33 & 0.922 & 0.107 & 4.382 & 0.611
& 25.60 & 0.866 & 0.165 & 4.218 & 0.583 \\
\makecell[l]{WF-Diff\\\citep{WF-Diff}} & CVPR'24
& 22.09 & 0.839 & 0.146 & 4.278 & 0.596
& 24.60 & 0.846 & 0.162 & 4.283 & 0.572 \\
\makecell[l]{SeaDiff\\\citep{SeaDiff}} & TCSVT'25
& 23.45 & 0.912 & 0.111 & 4.357 & \textbf{0.617}
& 26.82 & 0.849 & 0.131 & 4.238 & 0.586 \\
\makecell[l]{DiffColor\\\citep{DiffColor}} & TMM'26
& 21.93 & 0.836 & 0.153 & 4.440 & 0.592
& 27.10 & 0.861 & 0.147 & 4.231 & 0.587 \\
\makecell[l]{Reti-Diff\\\citep{Reti-Diff}} & ICLR'25
& 26.06 & 0.926 & 0.086 & 4.357 & 0.616
& 26.51 & 0.889 & 0.159 & 4.293 & \textbf{0.604} \\
\makecell[l]{WWE-UIE\\\citep{WWE-UIE}} & WACV'26
& 24.24 & 0.920 & 0.102 & 4.382 & 0.615
& 27.18 & 0.855 & 0.146 & 4.243 & 0.586 \\
\textbf{Ours} & -
& \textbf{26.85} & \textbf{0.931} & \textbf{0.081} & 4.359 & \textbf{0.617}
& \textbf{33.06} & \textbf{0.933} & \textbf{0.069} & 4.298 & 0.594 \\
\bottomrule
\end{tabular}
\end{adjustbox}
}
\end{table}

\subsubsection{Quantitative Comparisons}
Table~\ref{tab:main_results_uieb_lsui} reports the quantitative results on UIEB. RPL-UIE achieved the best PSNR, SSIM, LPIPS, and UCIQE values of 26.85 dB, 0.931, 0.081, and 0.617, respectively. Compared with Reti-Diff, which achieved the second-best PSNR, SSIM, and LPIPS results, RPL-UIE improved PSNR and SSIM by 0.79 dB and 0.005, respectively, while reducing LPIPS by 0.005. Although several methods obtained higher UIQM values, RPL-UIE performed better in terms of PSNR, SSIM, and LPIPS.

Table~\ref{tab:main_results_uieb_lsui} also reports the results on LSUI. RPL-UIE achieved the best PSNR, SSIM, and LPIPS values of 33.06 dB, 0.933, and 0.069, respectively. Relative to the second-best result for each metric, RPL-UIE improved PSNR by 5.88 dB over WWE-UIE and SSIM by 0.044 over Reti-Diff, while reducing LPIPS by 0.062 relative to SeaDiff.

\begin{table}[pos=!htbp]
\centering
\normalfont\rmfamily\fontsize{7.5}{8.4}\selectfont
\setlength{\tabcolsep}{1.0pt}
\renewcommand{\arraystretch}{1.03}

\EAAITable{Quantitative comparison on the EUVP-Dark, EUVP-ImageNet, and EUVP-Scenes datasets. Best results are in bold.}{tab:main_results_euvp}{%
\begin{adjustbox}{max width=\linewidth}
\begin{tabular}{@{}llccccccccccccccc@{}}
\toprule
Method & Venue
& \multicolumn{5}{c}{EUVP-Dark}
& \multicolumn{5}{c}{EUVP-ImageNet}
& \multicolumn{5}{c}{EUVP-Scenes} \\
\cmidrule(lr){3-7}\cmidrule(lr){8-12}\cmidrule(lr){13-17}
& & PSNR\textuparrow & SSIM\textuparrow & LPIPS\textdownarrow & UIQM\textuparrow & UCIQE\textuparrow
& PSNR\textuparrow & SSIM\textuparrow & LPIPS\textdownarrow & UIQM\textuparrow & UCIQE\textuparrow
& PSNR\textuparrow & SSIM\textuparrow & LPIPS\textdownarrow & UIQM\textuparrow & UCIQE\textuparrow \\
\midrule
\makecell[l]{U-shape\\\citep{U-shape}} & TIP'23
& 22.08 & 0.900 & 0.241 & 4.477 & 0.557
& 24.83 & 0.821 & 0.213 & 4.455 & 0.592
& 25.47 & 0.794 & 0.258 & 4.381 & 0.584 \\
\makecell[l]{CCL-Net\\\citep{CCL-Net}} & TMM'25
& 21.48 & 0.890 & 0.232 & 4.500 & 0.572
& 22.95 & 0.790 & 0.212 & 4.500 & \textbf{0.600}
& 24.04 & 0.793 & 0.156 & \textbf{4.443} & 0.584 \\
\makecell[l]{GuidedHybSensUIR\\\citep{GuidedHybSensUIR}} & TCSVT'25
& \textbf{22.60} & \textbf{0.912} & 0.245 & 4.452 & 0.537
& 24.79 & 0.834 & 0.198 & 4.431 & 0.591
& 25.55 & 0.839 & 0.176 & 4.345 & 0.582 \\
\makecell[l]{WF-Diff\\\citep{WF-Diff}} & CVPR'24
& 20.91 & 0.895 & 0.241 & 4.512 & \textbf{0.582}
& 24.05 & 0.822 & 0.162 & 4.513 & 0.596
& 26.87 & 0.860 & 0.134 & 4.437 & 0.586 \\
\makecell[l]{SeaDiff\\\citep{SeaDiff}} & TCSVT'25
& 21.56 & 0.900 & \textbf{0.216} & 4.468 & 0.560
& 23.67 & 0.800 & \textbf{0.142} & 4.531 & 0.583
& 26.76 & 0.824 & 0.118 & 4.357 & 0.581 \\
\makecell[l]{DiffColor\\\citep{DiffColor}} & TMM'26
& 21.65 & 0.903 & 0.239 & 4.418 & 0.569
& 25.09 & 0.829 & 0.165 & 4.440 & 0.592
& 26.78 & 0.839 & 0.176 & 4.317 & 0.584 \\
\makecell[l]{Reti-Diff\\\citep{Reti-Diff}} & ICLR'25
& 22.02 & 0.905 & 0.243 & 4.478 & 0.551
& 25.77 & 0.853 & 0.174 & \textbf{4.532} & 0.592
& 29.25 & 0.895 & 0.091 & 4.365 & \textbf{0.591} \\
\makecell[l]{WWE-UIE\\\citep{WWE-UIE}} & WACV'26
& 22.29 & 0.907 & 0.247 & 4.466 & 0.534
& 25.46 & 0.841 & 0.161 & 4.465 & 0.592
& 26.82 & 0.837 & 0.151 & 4.331 & 0.584 \\
\textbf{Ours} & -
& 22.52 & 0.910 & 0.225 & \textbf{4.513} & 0.551
& \textbf{26.58} & \textbf{0.872} & 0.146 & 4.490 & 0.598
& \textbf{29.39} & \textbf{0.896} & \textbf{0.085} & 4.366 & \textbf{0.591} \\
\bottomrule
\end{tabular}
\end{adjustbox}
}
\end{table}

Table~\ref{tab:main_results_euvp} reports the quantitative results on the three EUVP subsets. On EUVP-Dark, RPL-UIE obtained PSNR, SSIM, LPIPS, and UIQM values of 22.52 dB, 0.910, 0.225, and 4.513, respectively. It ranked second in PSNR, SSIM, and LPIPS and achieved the best UIQM. GuidedHybSensUIR obtained the best PSNR and SSIM values of 22.60 dB and 0.912, exceeding RPL-UIE by only 0.08 dB and 0.002, respectively. Meanwhile, RPL-UIE reduced LPIPS by 0.020 and increased UIQM by 0.061 relative to GuidedHybSensUIR. On EUVP-ImageNet, RPL-UIE achieved the best PSNR and SSIM values of 26.58 dB and 0.872, respectively, while ranking second in LPIPS and UCIQE with values of 0.146 and 0.598. Compared with CCL-Net, which achieved the best UCIQE, RPL-UIE improved PSNR and SSIM by 3.63 dB and 0.082 and reduced LPIPS by 0.066, with a UCIQE difference of only 0.002. On EUVP-Scenes, RPL-UIE achieved the best PSNR, SSIM, and LPIPS values of 29.39 dB, 0.896, and 0.085, respectively. Compared with Reti-Diff, it improved PSNR and SSIM by 0.14 dB and 0.001 while reducing LPIPS by 0.006. 

Across UIEB, LSUI, and the three EUVP subsets, RPL-UIE achieves competitive or superior performance in PSNR, SSIM, and LPIPS, demonstrating strong reconstruction fidelity, structural similarity, and perceptual consistency under diverse underwater degradations. The results across multiple benchmarks further suggest that the complementary appearance and photometric priors, together with teacher-supervised residual refinement and frequency-aware calibration, provide effective guidance for underwater image enhancement under different degradation patterns.

\begin{figure}[pos=!htbp]
\centering
\includegraphics[width=\textwidth]{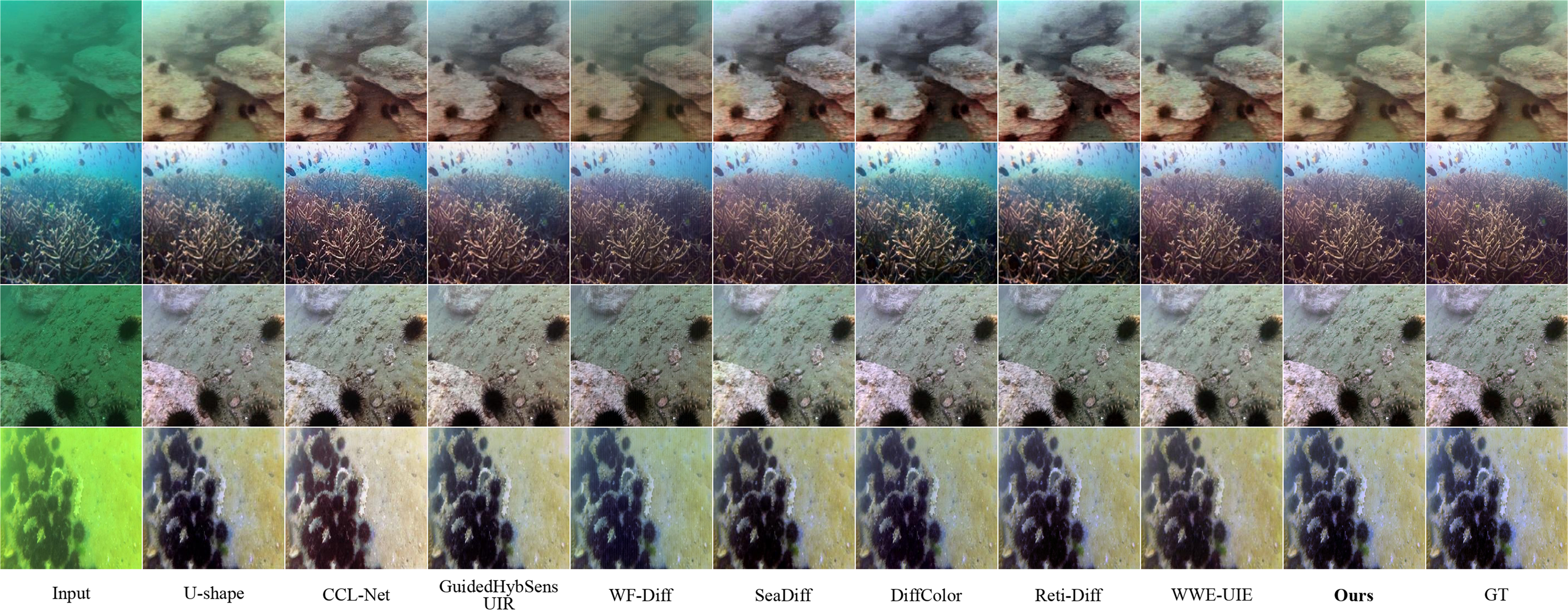}
\caption{Visual comparison on the LSUI dataset.}
\label{fig:lsui_visual}
\end{figure}

\subsubsection{Qualitative Comparisons}

Fig.~\ref{fig:visual_comparison} presents qualitative comparisons on UIEB. Most methods improve image visibility but differ in color fidelity and structural preservation. WF-Diff and Reti-Diff produce relatively strong color and contrast changes in several examples, whereas RPL-UIE maintains a more reference-consistent appearance while preserving object contours and local textures. Fig.~\ref{fig:lsui_visual} presents qualitative comparisons on LSUI. U-shape leaves some low-visibility regions insufficiently restored, while DiffColor exhibits scene-dependent inconsistencies in color and brightness recovery. In comparison, RPL-UIE achieves a better balance between illumination correction and detail preservation, particularly for coral structures, seafloor textures, and foreground objects.

\begin{figure}[pos=!htbp]
\centering
\includegraphics[width=\textwidth]{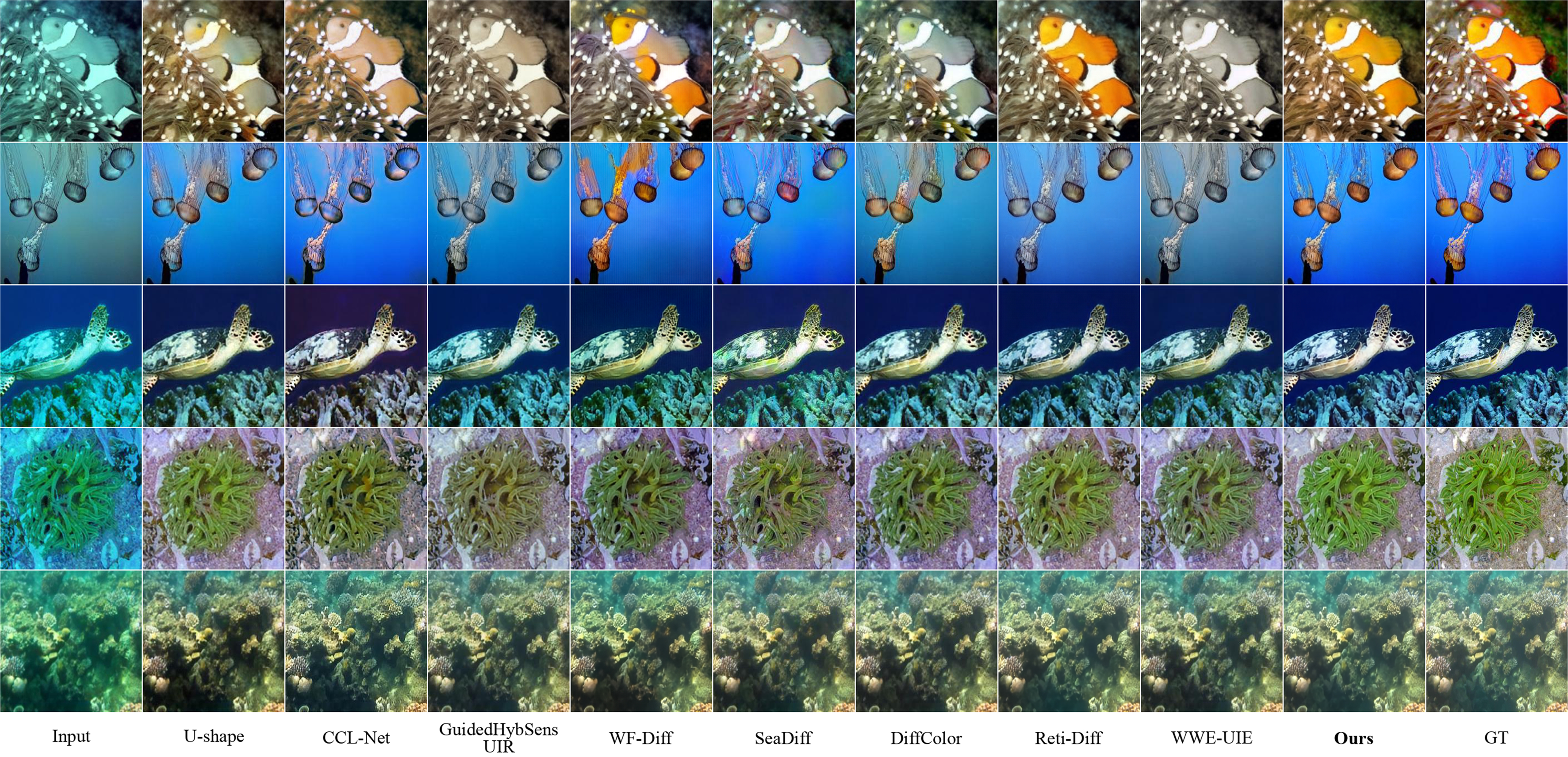}
\caption{Visual comparison on the EUVP dataset.}
\label{fig:euvp_visual}
\end{figure}

Fig.~\ref{fig:euvp_visual} presents qualitative comparisons on EUVP. GuidedHybSensUIR and WWE-UIE improve overall visibility but provide less complete foreground color restoration in the fish and jellyfish examples. In contrast, RPL-UIE achieves more natural color restoration, yielding color distributions closer to those of the reference images while preserving foreground structures and background details across diverse scenes. These qualitative results are consistent with the complementary appearance and photometric guidance provided by the dual priors, while prior refinement and calibration help reduce overcorrection and unstable detail responses.

\subsection{Real-world Scenario Validation}

To further assess the performance of RPL-UIE in real-world underwater scenes, we used an ROV platform to capture an underwater video in Dushu Lake and enhanced the recorded frames. The scene exhibited severe water turbidity, nonuniform illumination, and viewpoint changes caused by ROV motion. Object boundaries and background structures were indistinct in the raw frames, posing substantial challenges to subsequent inter-frame matching and visual perception.

\begin{figure}[pos=H]
\centering
\includegraphics[width=\columnwidth,height=0.84\textheight,keepaspectratio]{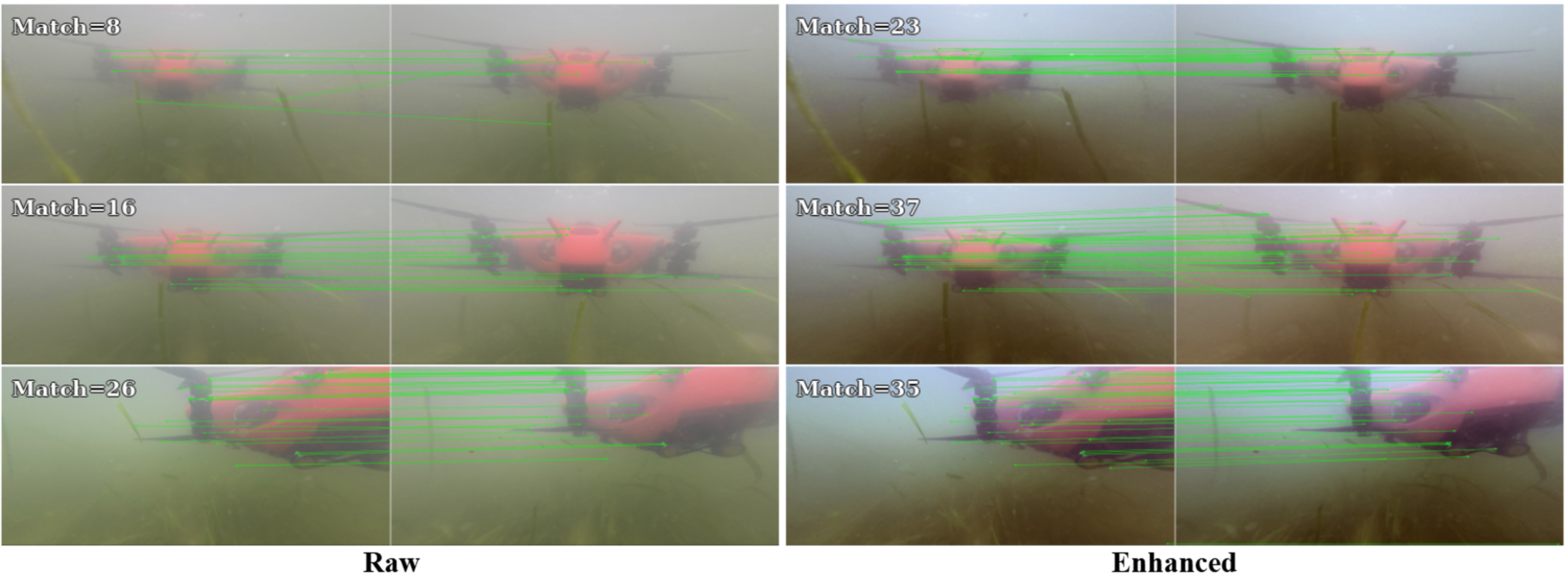}
\caption{SIFT feature matching between adjacent frames from a real-world underwater video. The left and right columns show the matching results for the raw and enhanced frames, respectively.}
\label{fig:real_scene_matching}
\end{figure}

Fig.~\ref{fig:real_scene_matching} compares SIFT matching between adjacent raw and enhanced video frames. We performed SIFT~\citep{SIFT} feature matching on both sets of frames and counted the valid matches. For the three displayed frame pairs, the raw frames yielded 8, 16, and 26 valid matches, whereas the enhanced frames yielded 23, 37, and 35, respectively. In the enhanced frames, the ROV body, frame edges, and surrounding underwater structures are more distinguishable, yielding more valid inter-frame matches.

These results indicate that RPL-UIE improves local structural visibility and inter-frame feature matching in the captured turbid underwater scene, which may benefit subsequent underwater visual navigation.

\EAAIkeepwith{8\baselineskip}
\subsection{Downstream Underwater Vision Tasks}
\subsubsection{Underwater Object Detection}

To assess the effect of enhancement on downstream underwater vision, we conducted object detection experiments on the UDD dataset~\citep{UDD}. UDD contains 1,827 training images and 400 test images covering three categories: holothurian, echinus, and scallop. For each enhancement method, we processed both the training and test sets to construct the corresponding detection dataset. A YOLOv9s detector~\citep{YOLOv9} was then retrained under identical settings for each dataset. All detectors were trained for 100 epochs with a batch size of 16.

\begin{table}[pos=!htbp]
\centering
\scriptsize
\setlength{\tabcolsep}{4.6pt}
\renewcommand{\arraystretch}{1.10}

\EAAITable{Object detection results of different enhancement methods on the UDD dataset.}{tab:downstream_detection}{%
\resizebox{\textwidth}{!}{%
\begin{tabular}{@{}lcccccccccc@{}}
\toprule
\multirow{2}{*}{Method}
& \multirow{2}{*}{P}
& \multirow{2}{*}{R}
& \multicolumn{3}{c}{AP$_{50}\uparrow$}
& \multirow{2}{*}{mAP$_{50}\uparrow$}
& \multicolumn{3}{c}{AP$_{50:95}\uparrow$}
& \multirow{2}{*}{mAP$_{50:95}\uparrow$} \\
\cmidrule(lr){4-6} \cmidrule(lr){8-10}
& & & Holothurian & Echinus & Scallop
& & Holothurian & Echinus & Scallop & \\
\midrule
Origin & 64.5 & 59.3 & 50.5 & \textbf{90.6} & 48.0 & 63.0 & 17.9 & 40.5 & 17.8 & 25.4 \\
\makecell[l]{CCL-Net\\\citep{CCL-Net}} & 69.1 & 55.1 & 49.6 & 89.8 & 43.6 & 61.0 & 20.4 & 40.7 & 17.1 & 26.1 \\
\makecell[l]{GuidedHybSensUIR\\\citep{GuidedHybSensUIR}} & \textbf{69.9} & 59.1 & 52.5 & 90.1 & 52.9 & 65.1 & 20.6 & 40.9 & 18.3 & 26.6 \\
\makecell[l]{WF-Diff\\\citep{WF-Diff}} & 65.9 & 60.9 & 51.3 & 90.4 & 50.0 & 63.9 & 20.1 & 41.1 & 17.7 & 26.3 \\
\makecell[l]{DiffColor\\\citep{DiffColor}} & 69.0 & 61.0 & 50.6 & 89.2 & \textbf{53.8} & 64.5 & 19.9 & 40.7 & \textbf{20.4} & 27.0 \\
\makecell[l]{WWE-UIE\\\citep{WWE-UIE}} & 62.8 & 60.7 & \textbf{53.1} & 89.4 & 47.4 & 63.3 & 20.6 & 41.1 & 18.3 & 26.7 \\
Ours & 64.5 & \textbf{64.1} & 52.8 & 90.3 & 53.1 & \textbf{65.4} & \textbf{21.1} & \textbf{41.5} & 19.6 & \textbf{27.4} \\
\bottomrule
\end{tabular}%
}
}
\end{table}

\begin{figure}[pos=!htbp]
\centering
\includegraphics[width=\textwidth]{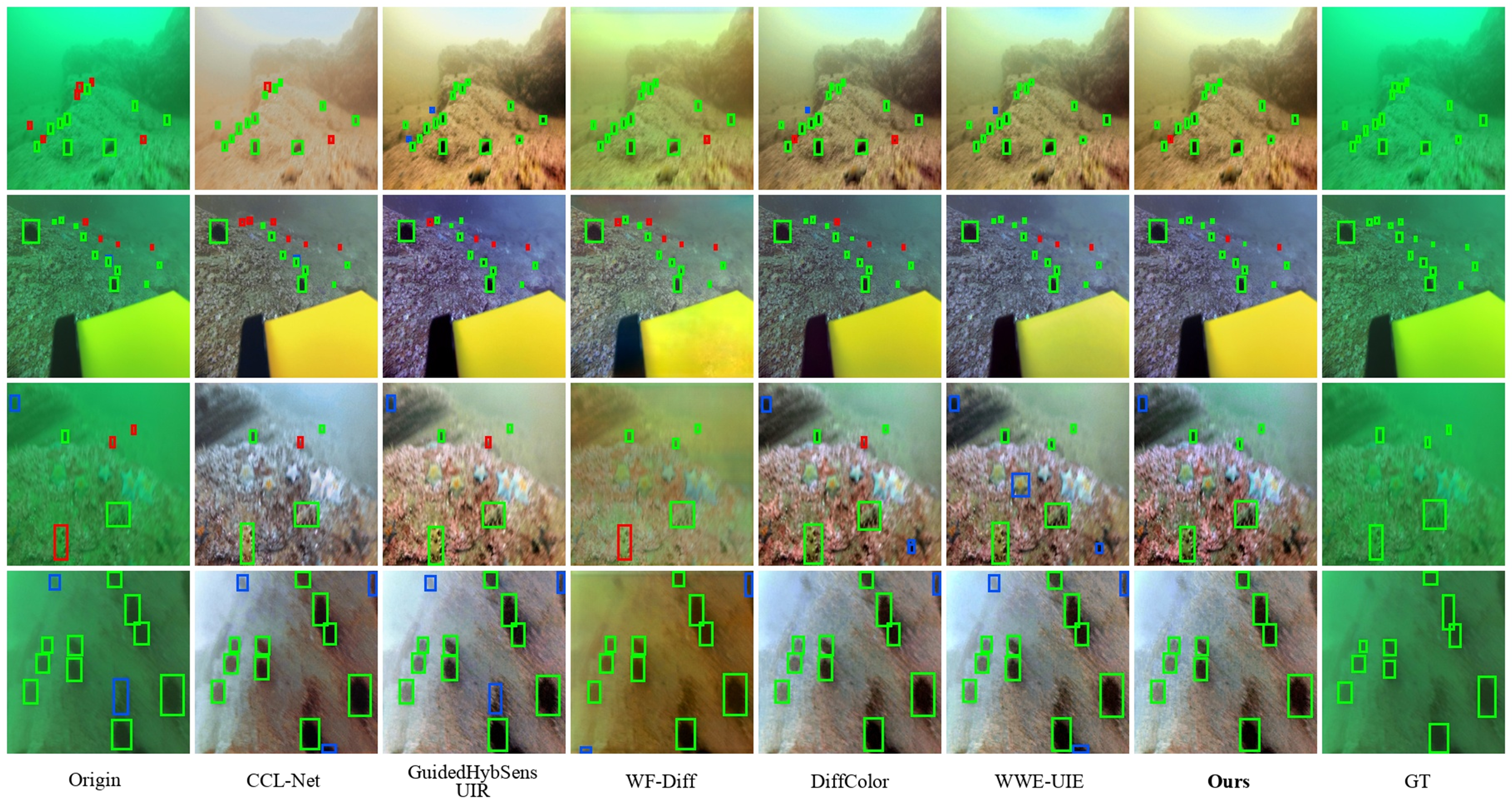}
\caption{Visual comparison of downstream object detection on UDD. Green, blue, and red boxes indicate correct detections, false positives, and missed objects, respectively.}
\label{fig:downstream_detection_visual}
\end{figure}

Table~\ref{tab:downstream_detection} shows that RPL-UIE increased mAP$_{50}$ and mAP$_{50:95}$ from 63.0 and 25.4 on the original images to 65.4 and 27.4, respectively, achieving the highest overall values among the evaluated enhancement methods. At the category level, it improved AP$_{50}$ for holothurian and scallop while remaining competitive for echinus. These gains support the utility of RPL-UIE-enhanced images for underwater object detection. Together with the qualitative examples, the results are consistent with improved preservation of object-level structures and local details.

Fig.~\ref{fig:downstream_detection_visual} further presents representative detection results. Although most enhancement methods alleviated the dominant color cast and improved object visibility, false positives and missed detections remained in complex backgrounds and low-contrast regions. In comparison, the RPL-UIE examples exhibit clearer object structures and local details together with more accurate detections after color correction.

\subsubsection{Underwater Instance Segmentation}

We further evaluated instance segmentation on UIIS~\citep{UIIS}, which contains 3,937 training images and 691 test images across seven categories. Following the detection protocol, each enhancement method was applied to both splits, and SOLOv2~\citep{SOLOv2} was retrained on the corresponding enhanced training set for 24 epochs with a batch size of 16.

\begin{table}[pos=!htbp]
\centering
\scriptsize
\setlength{\tabcolsep}{3.2pt}
\renewcommand{\arraystretch}{1.10}

\EAAITable{Instance segmentation results of different enhancement methods on the UIIS dataset.}{tab:downstream_segmentation_solov2_24ep}{%
\resizebox{\textwidth}{!}{%
\begin{tabular}{@{}lcccccccccccccccc@{}}
\toprule
\multirow{2}{*}{Method}
& \multicolumn{7}{c}{AP$_{50}\uparrow$}
& \multirow{2}{*}{mAP$_{50}\uparrow$}
& \multicolumn{7}{c}{AP$_{50:95}\uparrow$}
& \multirow{2}{*}{mAP$_{50:95}\uparrow$} \\
\cmidrule(lr){2-8} \cmidrule(lr){10-16}
& Fish & Reefs & Aqua. & Wrecks & Divers & Robots & Sea-floor
& & Fish & Reefs & Aqua. & Wrecks & Divers & Robots & Sea-floor & \\
\midrule
Origin & \textbf{64.5} & 46.2 & 12.5 & 18.0 & 79.0 & 4.0 & 13.7 & 34.0 & \textbf{39.8} & 28.9 & 5.8 & 14.1 & 49.4 & 3.0 & 8.7 & 21.4 \\
\makecell[l]{CCL-Net\\\citep{CCL-Net}} & 62.3 & 44.8 & 12.2 & \textbf{27.1} & 80.3 & 9.7 & 14.8 & 35.9 & 38.2 & 27.2 & 5.4 & 19.4 & 50.4 & 7.7 & 8.6 & 22.4 \\
\makecell[l]{GuidedHybSensUIR\\\citep{GuidedHybSensUIR}} & 62.9 & 45.8 & 13.6 & 20.4 & 81.2 & 10.2 & 15.4 & 35.7 & 38.7 & 28.5 & 6.0 & 16.0 & \textbf{52.1} & 7.8 & 9.4 & 22.7 \\
\makecell[l]{WF-Diff\\\citep{WF-Diff}} & 63.2 & 45.2 & 12.7 & 25.2 & 78.8 & 8.4 & 15.3 & 35.6 & 39.2 & 28.1 & 5.8 & \textbf{19.8} & 48.8 & 6.5 & 10.3 & 22.7 \\
\makecell[l]{DiffColor\\\citep{DiffColor}} & 63.0 & \textbf{46.5} & \textbf{14.2} & 19.4 & \textbf{81.9} & 5.3 & 15.6 & 35.1 & 38.9 & 28.7 & \textbf{6.4} & 14.9 & 51.4 & 3.9 & 9.6 & 22.0 \\
\makecell[l]{WWE-UIE\\\citep{WWE-UIE}} & 63.7 & 46.4 & 12.6 & 20.0 & 81.2 & 4.1 & \textbf{17.1} & 35.0 & 39.6 & \textbf{29.1} & 5.6 & 15.5 & 51.7 & 2.7 & \textbf{11.2} & 22.2 \\
Ours & 63.6 & 45.5 & 12.7 & 23.2 & \textbf{81.9} & \textbf{13.9} & 14.7 & \textbf{36.5} & 39.4 & 28.2 & 5.8 & 16.9 & 51.0 & \textbf{11.7} & 9.1 & \textbf{23.2} \\
\bottomrule
\end{tabular}%
}
}
\end{table}

\begin{figure}[pos=!htbp]
\centering
\includegraphics[width=\textwidth]{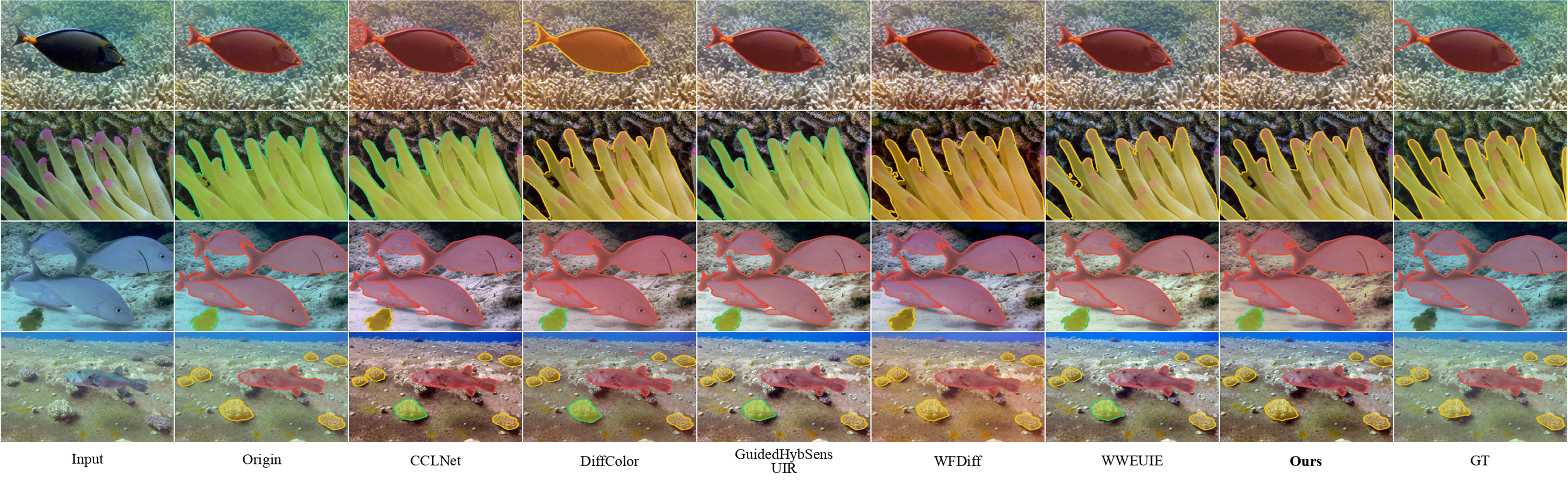}
\caption{Visual comparison of downstream instance segmentation on the UIIS dataset.}
\label{fig:downstream_segmentation_visual}
\end{figure}

Table~\ref{tab:downstream_segmentation_solov2_24ep} shows that RPL-UIE improved mAP$_{50}$ and mAP$_{50:95}$ from 34.0 and 21.4 on the original images to 36.5 and 23.2, achieving the best overall performance among the evaluated methods. It tied for the best AP$_{50}$ on human divers and achieved the best AP$_{50}$ and AP$_{50:95}$ on robots. These gains further support the utility of RPL-UIE-enhanced images for instance localization and mask prediction. The visual examples are also consistent with preservation of object contours and local details.
Fig.~\ref{fig:downstream_segmentation_visual} presents representative segmentation results on UIIS. Although most methods produce comparable masks for the primary instances, differences remain in boundary alignment, mask completeness, and small-region predictions. RPL-UIE maintains well-aligned object contours and competitive mask quality across the displayed examples.

\subsection{Ablation Study}
\subsubsection{Effectiveness of Main Components}
We conducted component ablations on UIEB, as summarized in Table~\ref{tab:component_ablation}. Starting from the reconstruction-only Baseline, configurations (A) and (B) introduce the appearance and photometric spatial prior branches, respectively, while configuration (C) combines both branches. Configuration (D) further incorporates RPRD for residual prior refinement, and the Full model additionally incorporates FPRC for prior residual calibration.

\begin{table}[pos=!htbp]
\centering
\footnotesize
\setlength{\tabcolsep}{3.0pt}
\renewcommand{\arraystretch}{1.08}

\EAAITable{Ablation analysis of the proposed components on the UIEB dataset.}{tab:component_ablation}{%
\begin{tabular}{@{}lccccccc@{}}
\toprule
Method & Appearance & Photometric & RPRD & FPRC
& PSNR$\uparrow$ & SSIM$\uparrow$ & LPIPS$\downarrow$ \\
\midrule
Baseline & $\times$ & $\times$ & $\times$ & $\times$ 
         & 25.03 & 0.925 & 0.092 \\
(A)      & $\surd$  & $\times$ & $\times$ & $\times$ 
         & 25.41 & 0.922 & 0.091 \\
(B)      & $\times$ & $\surd$  & $\times$ & $\times$ 
         & 25.78 & 0.924 & 0.088 \\
(C)      & $\surd$  & $\surd$  & $\times$ & $\times$ 
         & 26.20 & 0.925 & 0.085 \\
(D)      & $\surd$  & $\surd$  & $\surd$  & $\times$ 
         & 26.41 & 0.926 & 0.084 \\
\midrule
Full     & $\surd$  & $\surd$  & $\surd$  & $\surd$  
         & \textbf{26.85} & \textbf{0.931} & \textbf{0.081} \\
\bottomrule
\end{tabular}
}
\end{table}

\begin{figure}[pos=!htbp]
\centering
\includegraphics[width=\textwidth]{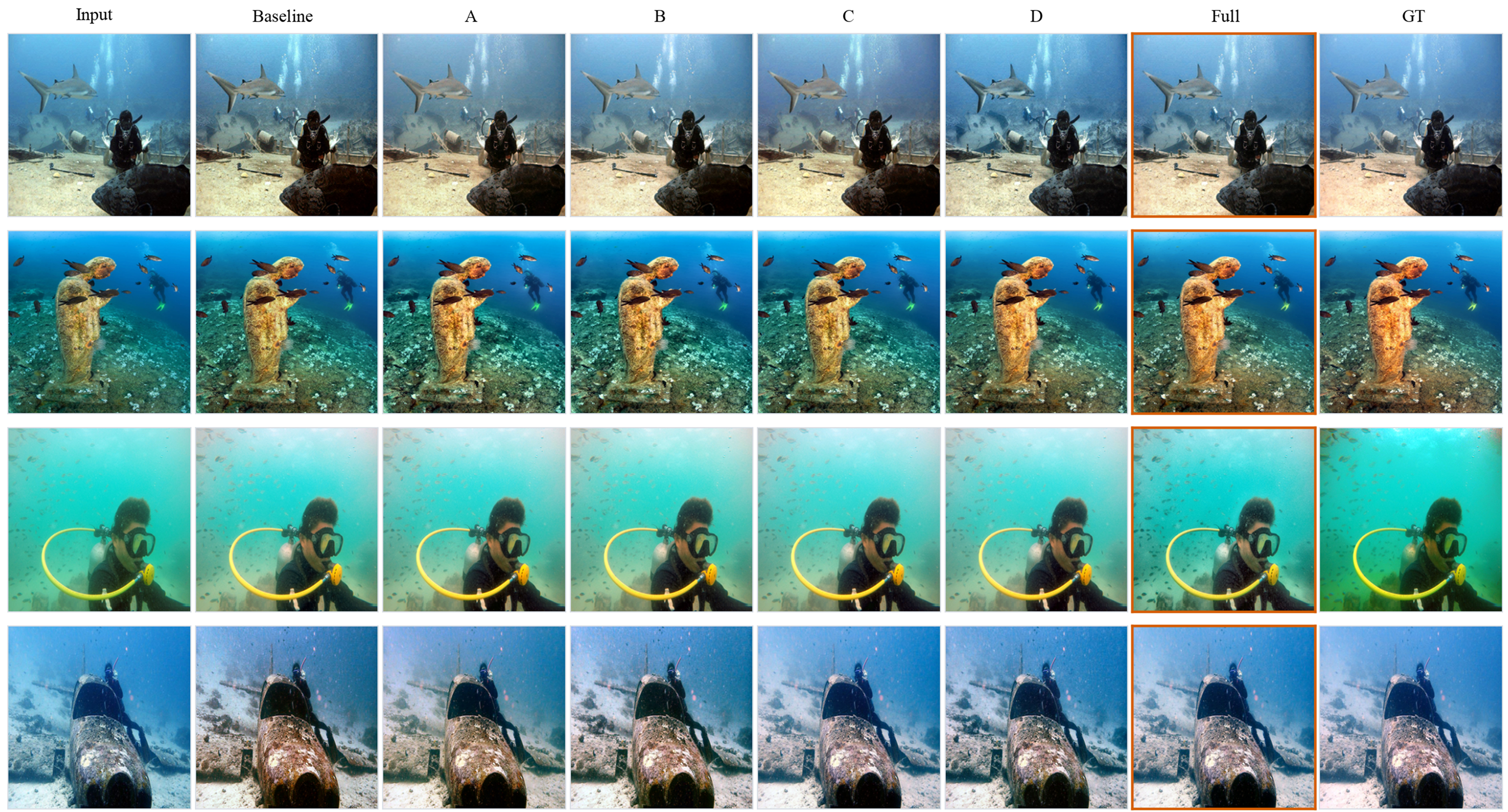}
\caption{Visual ablation of the main components in RPL-UIE.}
\label{fig:qualitative_ablation}
\end{figure}

Adding either the appearance or photometric prior improved PSNR and LPIPS, confirming the benefit of spatial prior guidance. The photometric prior yielded higher PSNR and lower LPIPS than the appearance prior, while combining both priors improved all three metrics, demonstrating their complementarity. Introducing RPRD yielded further gains, and FPRC enabled the Full model to achieve the best PSNR, SSIM, and LPIPS. 
Fig.~\ref{fig:qualitative_ablation} further illustrates the visual contribution of each component. The Baseline improves global brightness but retains residual color bias and insufficient foreground contrast in the shark and diver examples; the statue and shipwreck scenes also show differences from the reference images in object appearance and local contrast. Introducing either the appearance or photometric spatial prior improves color correction and foreground visibility, although the single-prior models still exhibit incomplete correction in some regions. Using both spatial priors provides a better balance between foreground and background restoration. Further introducing RPRD improves object appearance and local details, while the Full model with FPRC achieves the most balanced overall result in color consistency, foreground visibility, and detail preservation.

\subsubsection{Design Analysis of RPRD}

We evaluate three design choices in RPRD: whether forward diffusion is applied during training, whether noise is added to the deterministic initialization at inference, and whether the prediction target is the residual correction $\Delta\mathbf{P}^{b}$ or the complete teacher prior $\mathbf{P}^{b}_{t}$. In Table~\ref{tab:rprd_mechanism}, \emph{Train} indicates the use of forward diffusion during training, \emph{Start} denotes the deterministic initialization of the reverse process, and \emph{Noisy init.} indicates whether random noise is added to this initialization. Residual-target variants predict corrections conditioned on the coarse prior and start from the zero-residual state, whereas full-prior variants predict the complete teacher prior using the coarse prior as initialization. All subsequent reverse updates are deterministic.

\begin{table}[pos=!htbp]
\centering
\footnotesize
\setlength{\tabcolsep}{2.0pt}
\renewcommand{\arraystretch}{1.08}

\EAAITable{Design analysis of RPRD on UIEB.}{tab:rprd_mechanism}{%
\begin{adjustbox}{max width=\linewidth}
\begin{tabular}{@{}lcccccc@{}}
\toprule
Config & Target & Train & Start & Noisy init.
& PSNR$\uparrow$ & SSIM$\uparrow$ \\
\midrule
E1 & Residual & Yes & Zero-res. & Yes
& 26.50 & 0.929 \\
E2 & Residual & No  & Zero-res. & No
& 26.06 & 0.923 \\
E3 & Full prior & Yes & Coarse & Yes
& 26.28 & 0.925 \\
E4 & Full prior & Yes & Coarse & No
& 26.40 & 0.927 \\
\midrule
\textbf{Ours} & \textbf{Residual}
& \textbf{Yes} & \textbf{Zero-res.} & \textbf{No}
& \textbf{26.85} & \textbf{0.931} \\
\bottomrule
\end{tabular}
\end{adjustbox}
}
\end{table}

\begin{figure}[pos=!htbp]
\centering
\includegraphics[width=\columnwidth]{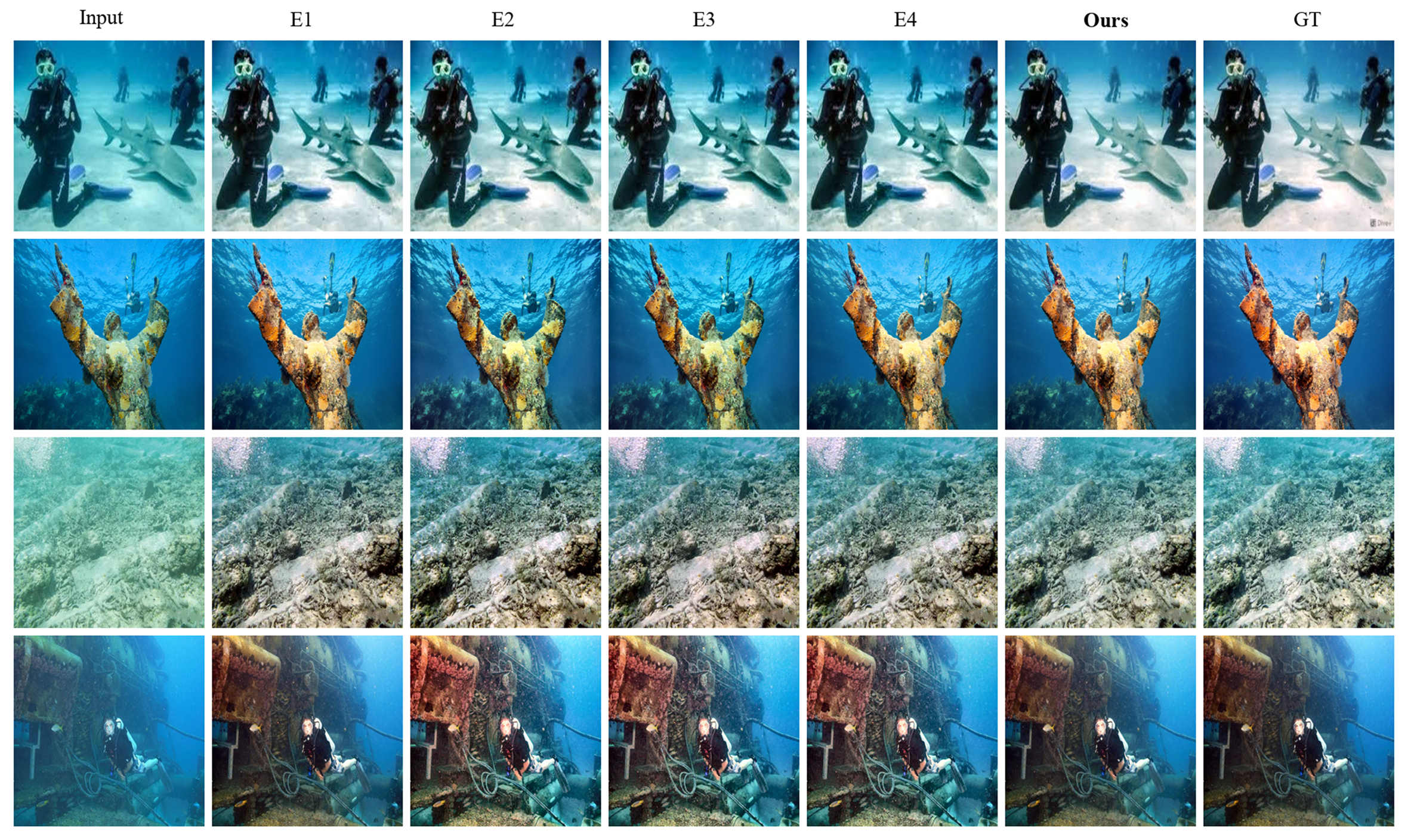}
\caption{Visual ablation of RPRD under different experimental configurations.}
\label{fig:rprd_design_visual}
\end{figure}

E2 differs from Ours only by omitting forward diffusion during training, and its lower PSNR and SSIM show that learning from noisy residual states contributes beyond multi-step reverse refinement. The E1/Ours and E3/E4 comparisons show that deterministic initialization outperforms noisy initialization for both residual-target and full-prior prediction, suggesting that added noise disrupts the input-dependent anchor provided by the coarse prior. Moreover, E1 outperforms E3 under noisy initialization, while Ours outperforms E4 under deterministic initialization, indicating that residual-target prediction is more effective than predicting the complete teacher prior. The visual comparisons in Fig.~\ref{fig:rprd_design_visual} further support these findings. Across the displayed scenes, Ours provides more balanced color and brightness recovery while preserving foreground structures and background details, producing results visually closer to the corresponding reference images.

\subsection{Limitations and Future Work}
Although RPL-UIE has been evaluated on multiple benchmarks and downstream vision tasks, its robustness under open-world underwater conditions, including extreme turbidity, severe low light, and substantial texture loss, warrants further investigation. In such cases, limited scene information in degraded inputs may reduce the accuracy of prior estimation and refinement in Stage II and ultimately affect reconstruction quality. Moreover, although diffusion modeling is confined to the prior residual space, RPRD still requires iterative reverse refinement and incurs additional inference overhead. Future work will explore larger and more diverse real-world underwater datasets, weakly supervised adaptation, and degradation-aware data construction to improve robustness, while investigating frequency-domain prior refinement with fewer denoising steps to achieve a better trade-off between restoration performance and computational efficiency.

\section{Conclusion}
We presented RPL-UIE, a two-stage teacher--student framework for reference-guided prior learning in content-preserving underwater image enhancement. In Stage I, the teacher network learns complementary reference-guided appearance and photometric priors from ground-truth information. These priors supervise Stage II, where the student network estimates the corresponding priors from degraded inputs alone, enabling reference-free inference.
To narrow the gap between the student-estimated priors and their teacher counterparts, RPRD progressively refines the coarse priors in the residual space, while FPRC calibrates the resulting prior residuals to guide image reconstruction.
Experiments on multiple UIE benchmarks demonstrated competitive enhancement performance, while ablation studies verified the effectiveness of the proposed components. Downstream detection, instance segmentation, and real-world ROV experiments further showed that RPL-UIE improves the practical utility of enhanced images for underwater visual perception.

\bibliographystyle{cas-model2-names}
\bibliography{references}
\end{document}